\documentclass{nature}

\usepackage{graphicx}
\usepackage{color}
\usepackage{comment}
\usepackage{subfig}
\usepackage{amsmath}
\usepackage{url}
							
\title{Weak ties strengthen anger contagion in social media}

\author{Rui Fan$^{1}$, Ke Xu$^1$ \& Jichang Zhao$^{2, 3, *}$}

\begin{document}

\maketitle

\begin{affiliations}
 \item State Key Lab of Software Development Environment, Beihang University
 \item School of Economics and Management, Beihang University
 \item Beijing Advanced Innovation Center for Big Data and Brain Computing\\
 $^*$ Corresponding author: jichang@buaa.edu.cn
\end{affiliations}

\begin{abstract}
Increasing evidence suggests that, similar to face-to-face communications, human emotions also spread in online social media. However, the mechanisms underlying this emotion contagion, for example, whether different feelings spread in unlikely ways or how the spread of emotions relates to the social network, is rarely investigated. Indeed, because of high costs and spatio-temporal limitations, explorations of this topic are challenging using conventional questionnaires or controlled experiments. Because they are collection points for natural affective responses of massive individuals, online social media sites offer an ideal proxy for tackling this issue from the perspective of computational social science. In this paper, based on the analysis of millions of tweets in Weibo, surprisingly, we find that anger travels easily along weaker ties than joy, meaning that it can infiltrate different communities and break free of local traps because strangers share such content more often. Through a simple diffusion model, we reveal that weaker ties speed up anger by applying both propagation velocity and coverage metrics. To the best of our knowledge, this is the first time that quantitative long-term evidence has been presented that reveals a difference in the mechanism by which joy and anger are disseminated. With the extensive proliferation of weak ties in booming social media, our results imply that the contagion of anger could be profoundly strengthened to globalize its negative impact.
\end{abstract}

Emotions have a substantial effect on human decision making and behavior~\cite{Gneezy28012014}. Emotions also transfer between different individuals through their communications and interactions, indicating that emotion contagion, which causes others to experience similar emotional states, could promote social interactions~\cite{Nummenmaa12062012,marsella2014computationally} and synchronize collective behavior, especially for individuals who are involved in social networks~\cite{Tadic2013} in the post-truth era~\cite{posttruth}. From this perspective, a better understanding of emotion contagion can disclose collective behavior patterns and help improve emotion management. However, the details of the mechanism by which emotion contagion spreads in a social network context remain unclear.

Conventional approaches such as laboratory experiments have been pervasively employed to attest to the existence of emotion contagion in real-world circumstances~\cite{hatfield1993emotional,fowler2008dynamic,rosenquist2011social}. However, unravelling the details of the mechanism behind emotion contagion is considerably more challenging because it is difficult for controlled experiments to establish large social networks, stimulate different emotions among the members of the network simultaneously, and then, track the propagation of emotions in real-time for long-term experiments. Meanwhile, to study the relationship dynamics between social structure and emotion contagion, properties such as the strength of relationships should also be considered; however, such considerations might introduce uncontrolled contextual factors that fundamentally undermine the reliability of the experiment. We argue that it is extremely difficult for conventional approaches to investigate the mechanism by which emotion spreads in the context of large social networks and long-term observations and that the traces of natural affective responses from the massive numbers of individuals in online social media offer a new and computationally suitable perspective~\cite{Lazer721,Golder1878,marsella2014computationally}.

It is not easy to differentiate between online interactions and face-to-face communications in terms of emotion contagion~\cite{cmc_emotion}. However, increasing evidence from both Facebook and Twitter in recent years has consistently demonstrated the existence of emotion contagion in online social media~\cite{guillory2011upset,coviello2014detecting,gruzd2011happiness,chmiel2011collective,kramer2012spread,dang2012impact,kramer2014experimental}, and the digital online media even upregulate user emotions to increase user engagement~\cite{GOLDENBERG2020316}. Kramer et al. provided the first experimental evidence of emotion contagion in Facebook by manipulating the amount of emotional content in people's News Feeds~\cite{kramer2014experimental}. Later Ferrara and Yang revealed evidence of emotion contagion in Twitter~\cite{ferrara2015measuring}; however, instead of controlling the content, they measured the emotional valence of user-generated content and showed that posting positive and negative tweets both follow paths that garner significant overexposure, indicating the spread of different feelings. Evidence from both studies suggests that---even in the absence of the non-verbal cues inherent to face-to-face interactions~\cite{ferrara2015measuring}, emotions that involve both positive and negative feelings can still be transferred from one user to others in online social media. In fact, through posting, reposting and other virtual interactions, users in online social media express and expose their natural emotional states into the social networks in real-time as the context evolves. Hence, serving as a ubiquitous sensing platform, online social media collects emotional expression and captures its dissemination in the most realistic and comprehensive circumstances, thus providing us with an unprecedented proxy for obtaining universal insights into the underlying mechanisms of emotion contagion in social networks.

However, in existing studies, emotions in online social media are usually simplified into either positive or negative~\cite{bollen2011happiness,bliss2012twitter,kramer2014experimental,ferrara2015measuring}, while many fine-grained emotional states---particularly negative ones such as anger and disgust---are neglected. In fact, on the Internet, negative feelings such as anger might dominate online bursts of communications about societal issues or terrorist attacks and play essential roles in driving collective behavior during the event propagation~\cite{chmiel2011negative,Zhao2012,fan2014anger}. Aiming to fill this critical gap, in this paper, we categorize human emotions into four categories~\cite{Zhao2012,fan2014anger}, joy, anger, disgust and sadness, and then attempt to disclose the mechanisms underlying their spread. Except for the type of emotion, contagion also tightly related to the underlying network structure such as connection strength~\cite{LIN2014342}. Previous study suggests that stronger ties lead to stronger emotion contagion~\cite{LIN201529}, but the relationship between the type of emotion and connection strength is still unrevealed. In this paper, we define the strength of connections on a large social media and study the relationship between tie strength and emotion contagion. Investigating fine-grained emotional states and the influence of tie strength enrich the landscape of emotion contagion and makes it possible to systematically understand the mechanisms that underlie the relationships between social structures and propagation dynamics.

\section*{Results}
\label{sec:res}

For this study, we collected 11,753,609 tweets posted by 92,176 users from September 2014 to March 2015, including the following networks of these users. Through a Bayesian classifier trained in~\cite{Zhao2012}, each emotional tweet in this data set was automatically classified as expressing joy, anger, disgust or sadness. However, we focus only on anger and joy in this paper since the possibility of contagion for disgust and sadness is trivial~\cite{fan2014anger}. Then, the structure preferences are accordingly investigated in the social network, which comprises approximately 100,000 subjects. Finally, a toy model were employed to investigate the conjecture that weak ties speed up anger in social media.~\footnote{All the data sets are publicly available at \url{https://dx.doi.org/10.6084/m9.figshare.4311920}.}.

\subsection{Anger prefers weaker ties than joy}
\label{subsec:ties}

The dynamics of emotion contagion are essentially coupled with the underlying social network, which provides channels for disseminating a sentiment from one individual to others. The key factor that determines users' actions is the relationships through which contagion is more likely to occur. Specifically, being able to predict which friend would be most likely to retweet an emotional post in the future would be a key insight for contagion modeling and control. 

Social relationships, or ties in social network, can be measured by their strength. Such measurements are instrumental in spreading both online and real-world human behaviors~\cite{Bond2012}. The strength of a tie in online social media can be intuitively measured using online interactions (i.e., retweets) between its two ends. Here, we present three metrics to depict tie strengths quantitatively. The first metric is the proportion of common friends~\cite{onnela2007structure,Zhao2010}, which is defined as $c_{ij}/(k_i-1+k_j-1-c_{ij})$ for the tie between users $i$ and $j$, where $c_{ij}$ denotes the number of friends that $i$ and $j$ have in common, and $k_i$ and $k_j$ represent the degrees of $i$ and $j$, respectively. Note that for the metric of common friends, the social network of Weibo is converted to an undirected network where each link represents possible interactions between both ends. The second metric is inspired by reciprocity in Twitter-like services: a higher ratio of reciprocity indicates more trust and more significant homophily~\cite{zhu2014influence}. Hence, for a given pair of users, the proportion of reciprocal retweets in their total flux is defined as the tie strength. The third metric is the number of retweets between two ends of a tie in Weibo, called retweet strength: larger values represent more frequent interactions. Note that, different from the previous two metrics, retweet strength evolves over time; therefore, here, we count only the retweets that occurred before the relevant emotional retweet. Moreover, to smooth the comparison between anger and joy, the retweet strength (denoted as $S$) is normalized by $(S-S_{min})/(S_{max}-S_{min})$, in which $S_{min}$ and $S_{max}$ separately represent the minimum and maximum values of all observations.

\begin{figure}
\centering
\includegraphics[width=0.6\linewidth]{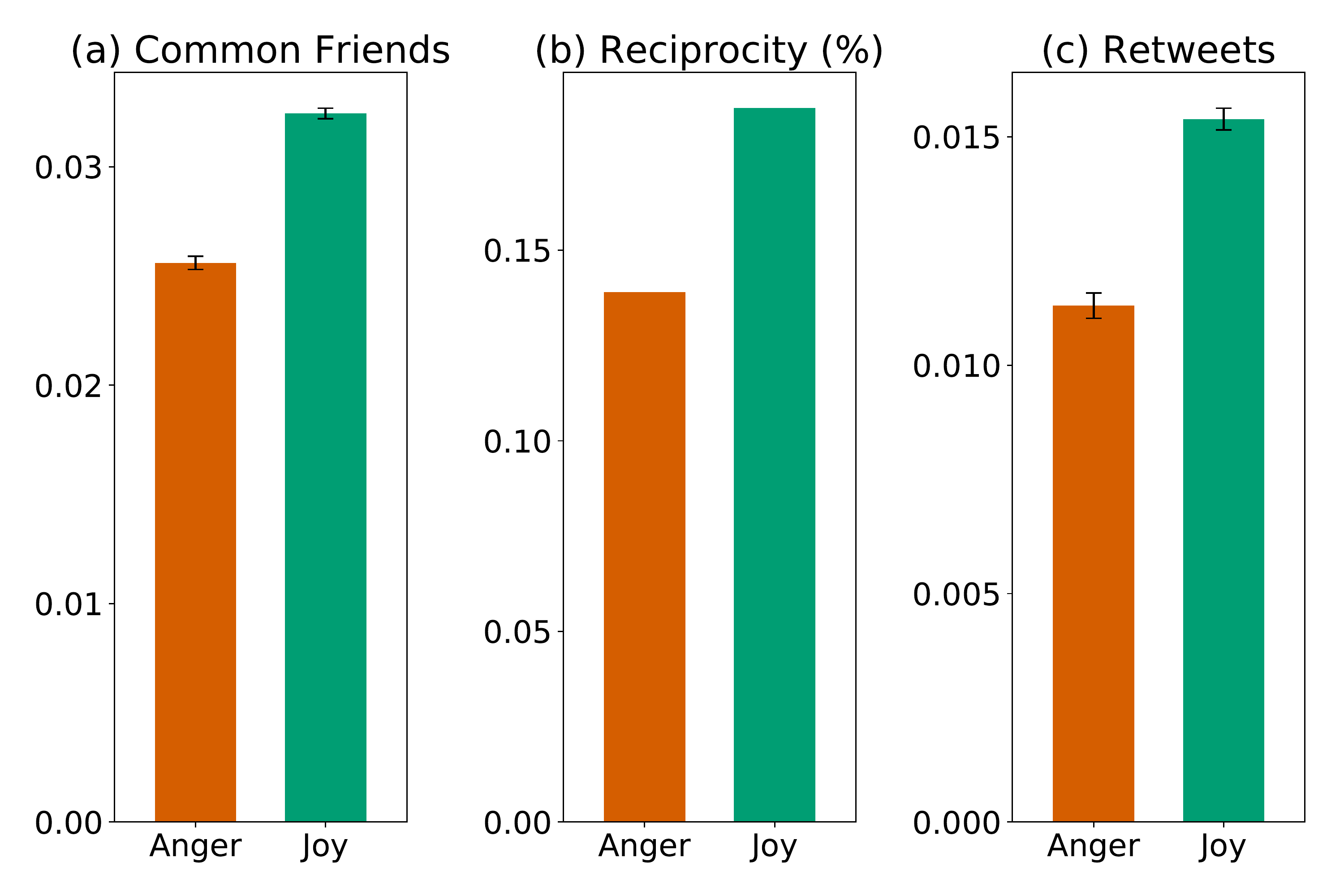}
\caption{Anger prefers weaker ties than joy. Three different metrics are averaged over all the emotional retweets in the dataset. The lower metrics for anger suggest that, in contagion, anger disseminates through weaker ties than joy. The error bars represent standard errors in (a) and (c), while in (b) there is no standard error because the reciprocity is simply a ratio obtained from all emotional retweets. All the three $p$-values calculated by Welch's two-sample $t$-test are almost zero, indicating the difference of means are prominent on all metrics.}
\label{fig:weak_tie}
\end{figure}

\begin{figure*}
\centering
\includegraphics[width=0.8\linewidth]{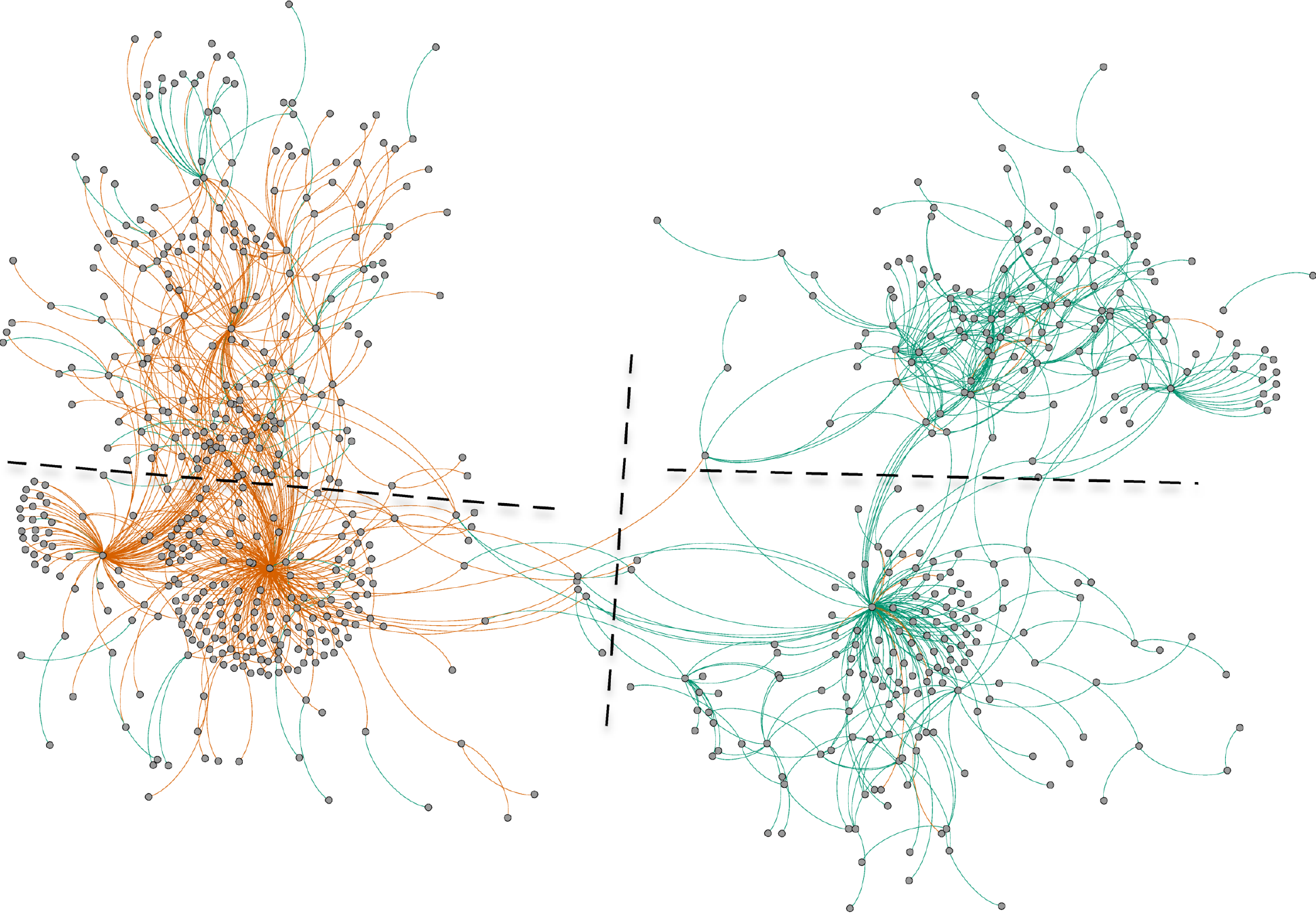}
\caption{Emotional contagion in a sampled snapshot of the Weibo network with four communities. Links with more angry retweets are orange; otherwise, they are green. It can be seen that anger prefers more weak ties that bridge different communities than does joy. The two communities to the left are dominated by anger; thus, a large percentage of their messages disseminate between communities. In contrast, the emotion of the two joy-dominated communities on the right tends to diffuse within the communities.}
\label{fig:graph}
\end{figure*}

By investigating each emotional retweet (i.e., a repost of an angry or joyful tweet posted by a followee), we can correlate emotion contagion with tie strengths. As shown in Fig.~\ref{fig:weak_tie}, all the metrics consistently demonstrate that anger prefers weaker ties in contagion than does joy, suggesting that angry tweets spread through weak ties with greater odds than do joyful tweets. It is well known that weak ties play essential roles in diffusion in social networks~\cite{granovetter1973strength}, particularly in breaking out of local traps caused by insular communities by bridging different clusters~\cite{onnela2007structure,Zhao2012KAIS,Meo2014}. For example, a typical snapshot of emotion contagion with four communities is illustrated in Fig.~\ref{fig:graph}, in which anger disseminates through weak ties of inter-communities more often than does joy. Because of this, compared to joy, anger has more chances to infiltrate different communities during emotion contagion because of its preference for weak ties. The increased number of infected communities leads to more global coverage, indicating that anger can achieve broader dissemination than joy over a given time period. 

The evidence from relationship strength suggests that weak ties make anger spread faster than joy, because it will infiltrate more communities in the Weibo social network. Next, a simple dissemination model in Weibo will be presented to provide further systematic support for this conjecture.

\subsection{Weak ties speed up anger in social media}
\label{subsec:fast}

No moods are created equally online~\cite{Choudhury2012}; the differences in contagion of positive and negative feelings has been a trending---but controversial---topic for years. Berger and Milkman revealed that positive content is more likely to become viral than negative content in a study of the NY Times~\cite{Berger2010}. Tadić and \v{S}uvakov found that, for human-like bots in online social networks, positive emotion bots are more effective than negative ones~\cite{Tadic2013}. Wu et al. even claimed that bad news containing more negative words fades more rapidly in Twitter~\cite{Wu2011}. In contrast, Chmiel et al. pointed out that negative sentiments boost user activity in BBC forums~\cite{chmiel2011negative} and Pfitzner et al. stated that users tend to retweet tweets with high emotional diversity~\cite{pfitzner2012emotional}. Ferrara and Yang demonstrated that although people are more likely to share positive content in Twitter, negative messages spread faster than positive ones at the content level~\cite{Ferra2015PJ}. Thus, Hansen et al. concluded that the relationship between emotion and virality is quite complicated~\cite{Hansen2011}. Here, we argue that a finer-grained division of human emotion, especially among negative emotions, will enrich the background for investigating contagion differences. In the meantime, creating explicit definitions of what fast spread entails will further facilitate the elimination of debate on this issue. 

To comprehensively understand how contagion tendency and tie strength function in the dynamics of emotion spread, we first establish a toy model to simulate such diffusion in Weibo's undirected following graph. In this model, which is inspired by the classic Susceptible-Infected (SI) model, first, a random seed with a certain emotion is selected to ignite the spread, and then, other susceptible nodes with no emotion will become infected and acquire the same sentimental state of the seed. Any susceptible node $s$ with an infected neighbor $i$ will become infected with the probability $p=\gamma~w_{is}^{\alpha}/{\sum_n{w_{in}^{\alpha}}}$, where $\gamma \ge 0$ reflects the contagious tendency, $\alpha$ controls the relationship strength preference and $n$ is one of $i$'s neighbors. Specifically, a greater $\gamma$ suggests that the emotion is more contagious, an $\alpha <0$ will preferentially select weak ties to spread the emotion, while strong ties are more likely to be selected when $\alpha > 0$, and $\alpha=0$ will cause a random selection of the diffusion path.

\begin{figure*}
\centering
\includegraphics[width=0.8\linewidth]{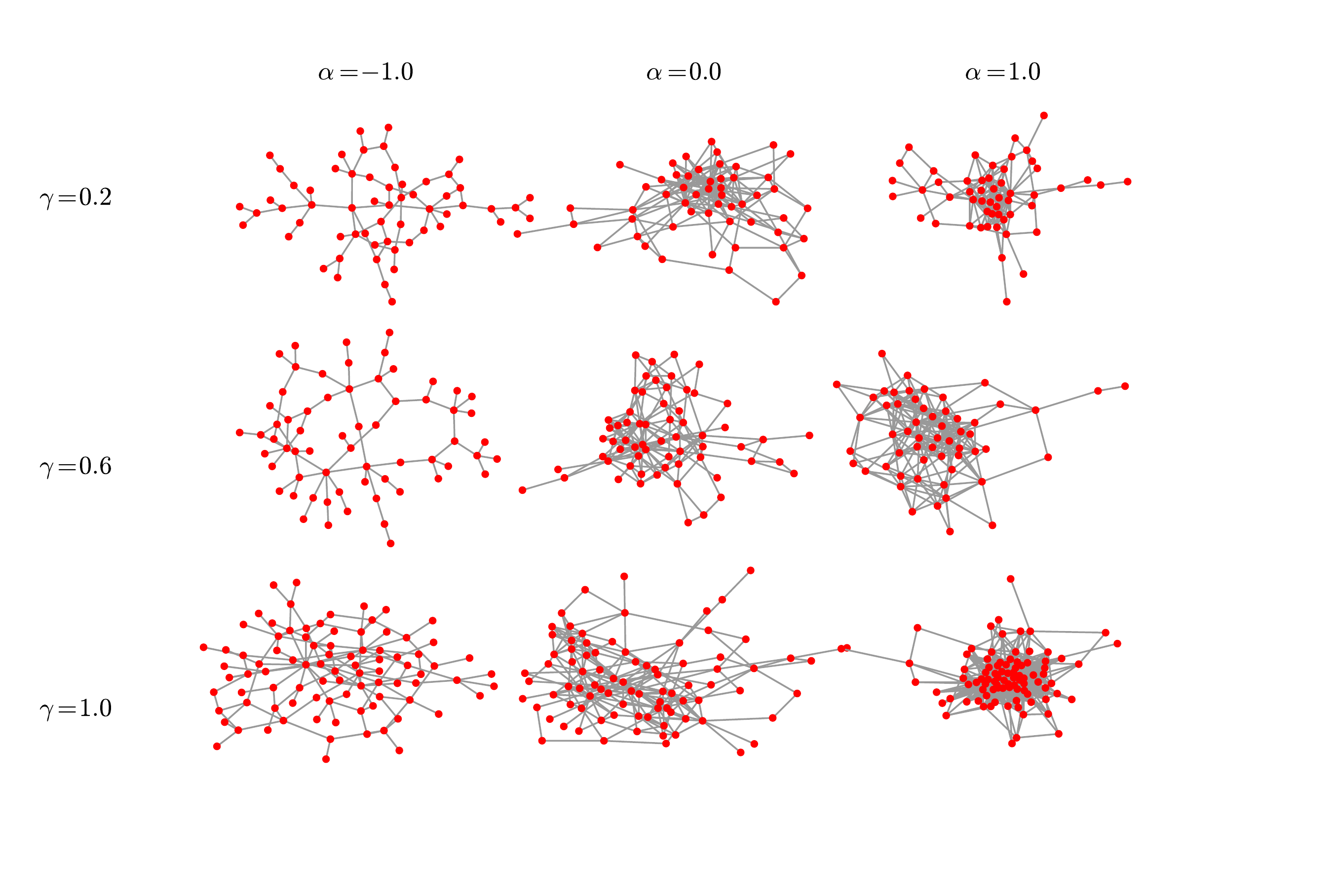}
\caption{Snapshots of the first infected 50 nodes in Weibo. Starting from the same seed, diverse values of $\gamma$ and $\alpha$ are used to simulate the spread. Note that in these simulations, the Weibo following graph was converted to an undirected graph.}
\label{fig:toy_model_snapshots}
\end{figure*}

As can be seen in Fig.~\ref{fig:toy_model_snapshots}, for a fixed $\gamma$, the infected networks produced by $\alpha=-1$ possess larger diameters than is the case for $\alpha=1$, and for a fixed $\alpha$, a greater $\gamma$ always leads to a more dense set of connections locally. Consistent with our previous conjecture, the preliminary observations above promisingly imply that, compared to strong ties, adopting weak ties as the diffusion paths improve the chances that an emotion will penetrate other network components, which then results in a large diameter and spare structures. In the meantime, a greater $\gamma$ means that more neighbors will be infected, which results in a dense neighborhood. By explicitly defining the speed and coverage of emotion spread, we next try to disclose which configurations of $\alpha$ and $\gamma$ will speed up propagation.

\begin{figure}
\centering
\includegraphics[width=0.9\linewidth]{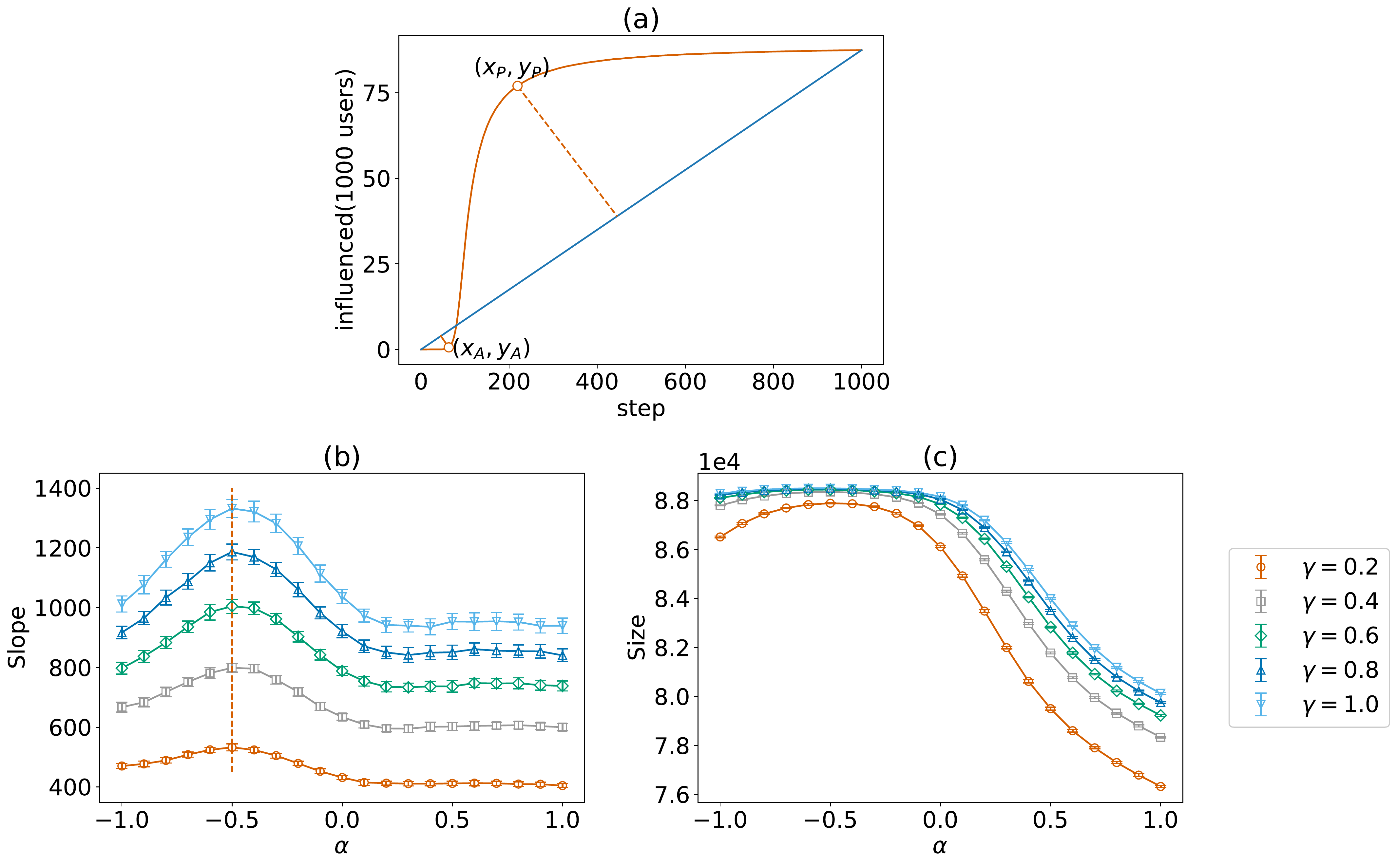}
\caption{Spread velocities and structures of the model. (a) The diagrammatic volume of infected nodes grows with time, where $x_A$ denotes the instant of awakening, $x_P$ denotes the peak time and $y_A$ and $y_P$ stand for the volume of infected nodes at the instants of awakening and peak, respectively. The first and last points of the accumulating curve are connected to obtain the reference line $L$. Then, the peak instant can be identified as the point the furthest distance from $L$ and located above $L$. The awakening instant can be found using the same method, except that the point will be located under $L$. (b) The slope varies with $\alpha$ for different $\gamma$ values. (c) The propagation coverage varies with $\alpha$ for different $\gamma$ values. All the simulations were performed on the undirected following graph of Weibo. For each pair of $\alpha$ and $\gamma$ values, 50 simulations with random seeds were conducted to guarantee stable statistics. The standard deviations are also presented.}
\label{fig:toy_model}
\end{figure}

For online emotion spread, ``fast spread'' means that the volume of emotional tweets grows quickly during the period from the awakening instant to the diffusion peak. Through a parameterless approach presented in~\cite{ke2015defining}, we can precisely locate the instants of both awakening and peak. Then, the spread speed is reflected by the average velocity of the growth in diffusion from awakening to peak. Specifically, as demonstrated in Fig.~\ref{fig:toy_model}(a), the average velocity can be defined as the slope (i.e., $(y_P-y_A)/(x_P-x_A)$). Besides, propagation coverage is intuitively defined as the number of finally infected users to reflect the scale of involved users.

As shown in Fig.~\ref{fig:toy_model}(c), the speed of emotion spread from the perspective of both velocity reaches its maximum as $\alpha \approx -0.5$, and larger $\gamma$ values generate higher maximum speed values. It can be concluded that when $\alpha<0$, preferentially selecting weak ties as the diffusion path can greatly boost the velocity of the spread. Meanwhile, the propagation sizes are stable when $\alpha<0$ but rapidly decline as $\alpha$ grows from 0 to 1, as shown in Fig.~\ref{fig:toy_model}(d). Recall the spread snapshots from Fig.~\ref{fig:toy_model_snapshots}. Here, an $\alpha<0$ is inclined to select weak ties that help emotion penetrate the more distant parts of the network, leading to a larger propagation coverage. Hence weak ties will simultaneously accelerate diffusion speed and enlarge propagation coverage. 

\begin{figure}
\centering
\includegraphics[width=0.9\linewidth]{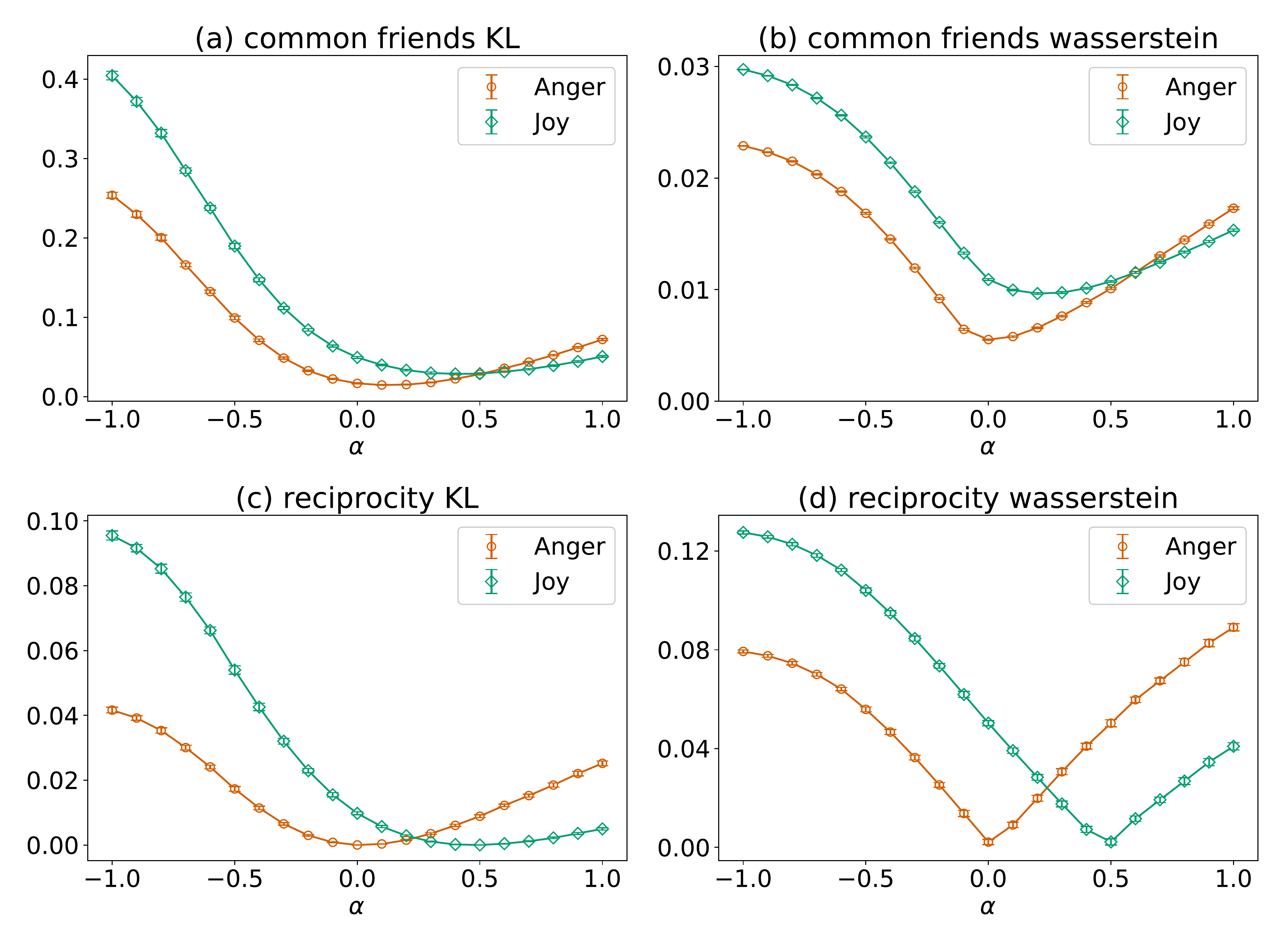}
\caption{KL and Wasserstein distance of contagion strength between model results and empirical data. (a) KL distance of anger and joy emotions based on common friends definition. (b) Wasserstein distance of anger and joy emotions based on common friends definition. (c) KL distance of anger and joy emotions based on reciprocity definition. (d) Wasserstein distance of anger and joy emotions based on reciprocity definition. $\gamma$ is fixed to 0.6 for simulation model.}
\label{fig:stat-test}
\end{figure}

To reveal the potential configuration of $\alpha$ and $\gamma$ in the model for anger and joy emotion, Kullback-–Leibler (KL) and Wasserstein divergences are calculated respectively on the tie-strength distribution between empirical data and model results. For common-friends measurement, the first distribution, denoted as $P(x)$, indicates the empirical strength distribution of all ties that angry (or joyful) retweets occur. The second distribution, denoted as $Q(x)$, means the common-friends strength distribution of ties through which diffusion occurs in simulation with specific $\alpha$ and $\gamma$. Then KL and Wasserstein divergences are calculated for $P(x)$ and $Q(x)$, representing the distance between anger (or joy) and the model with specific $\alpha$ and $\gamma$. For reciprocity measurement, $P(x)$ and $Q(x)$ are calculated similarly and both of them are Bernoulli distributions. Retweets measurement is not applied here because diffusion occur only once on one connection in the model. For both divergences, lower values indicate that the two distributions are more close. As can be seen in Fig.~\ref{fig:stat-test}, for both anger and joy, divergences decline with the growth of $\alpha$ when $\alpha$ is small. After achieving the valley, divergence values increase with the rising of $\alpha$. $\gamma$ is fixed to 0.6 in Fig.~\ref{fig:stat-test} because curves with different $\gamma$ are nearly overlapped, indicating that it has no influence for the conclusion (as seen in Fig.~\ref{fig:toy_model}(c)). It is worth noting that the minimum divergences for anger are always lower than that for joy emotion. For example, for KL divergences based on common friends measurement, the minimum value appear at $\alpha=0.1$ and $\alpha=0.4$ for anger and joy respectively. In other words, joy emotion prefer to diffuse between close friends ($\alpha=0.3\sim0.5$) but anger emotion diffuse randomly or only slightly prefer to spread between friends ($\alpha=0$ or $\alpha=0.1$). The fitted values of $\alpha$ for both anger and joy from realistic data suggest that emotion contagion is more likely to happen on ties with higher strengths, which is consistent with the previous study~\cite{LIN201529}. While the lower $\alpha$ of anger than that of joy further implies anger's preference on weaker ties, meaning odds of ties with lower strength to be involved in the spread of anger are unexpectedly greater. As seen in Fig.~\ref{fig:toy_model}, both diffusion velocity and coverage decline with the growth of $\alpha$ when $\alpha\geq0$, indicating that the mechanism of weaker ties preference speed up the diffusion of anger emotion.

To summarize, the presented toy model clearly discloses the roles of weaker ties in emotion spread, revealing that it helps emotion reach more susceptible individuals and increases the infection rate. The statistical divergence calculations further reveal that both anger and joy emotions are more likely to diffuse between friends but compared to joy, anger prefers weaker ties, which accelerate its diffusion speed and enlarge its propagation coverage. Weak ties were conventionally thought to act as bridges for novel information~\cite{granovetter1973strength} but our findings indicate that they will speed up the diffusion of anger in online social media. Additional evidence from real-world emotion spread is presented in the next section to support our explanations. 

\begin{figure}
\centering
\includegraphics[width=0.9\linewidth]{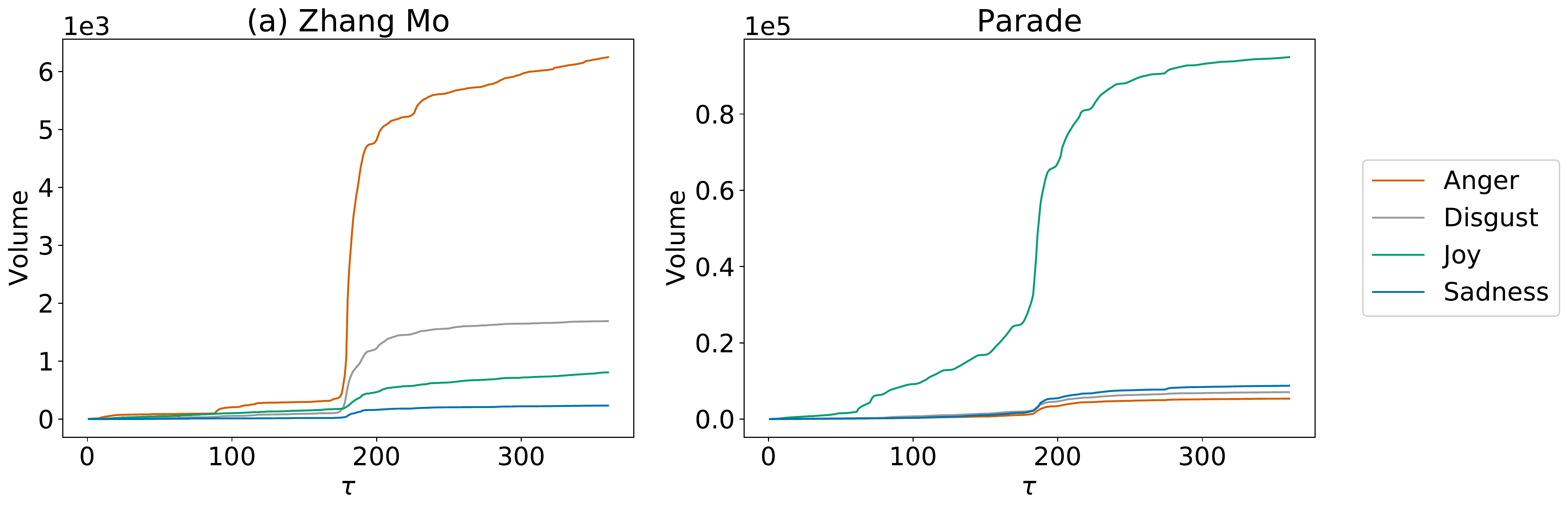}
\caption{Evidence from realistic online communication bursts. (a) An example of an anger-dominated event (celebrity scandal) in which cumulatively posted emotional tweets vary over time $\tau$. (b) An example of a joy-dominated event (a National Holiday parade).}
\label{fig:event}
\end{figure}

To further illustrate the promotions from weak ties on anger spread, over 600 communication burst events were extracted from Weibo. For each event, the emotion that occupied more than 60\% emotional tweets was defined as the dominant emotion of the event. In total we obtained 37 anger-dominated events and 200 joy-dominated events, as shown in Fig.~\ref{fig:event}(a), in which the cumulative volume of emotional tweets grows and then becomes saturated with time, at which point anger takes over the majority of the tweets. Here, the time granularity of all burst events is one hour. Several events with no obvious awakening or peak instants were omitted. It is worth noting that the small number of anger-dominated events explains its low prevalence in Weibo~\cite{Zhao2012}. The average results of spread velocity, i.e., the slope, are calculated and normalized by dividing the peak volume ($y_P$) of events. For anger- and joy-dominated events, the results are 0.027 and 0.020 respectively, which significantly support the simulation results that anger-dominated events tend to spread faster from the awakening to the peak than do joy-dominated events. We can conclude, based on evidence from both model simulations and real-world events, that anger indeed spreads faster than joy in social media such as Weibo.

\section*{Discussion}
\label{sec:dis}

It has been said that bad is always stronger than good~\cite{bad_is_stronger}. Emotional intensity decreases after explicit emotion expression in online social network, and this effect is stronger in negative than in positive emotions~\cite{fan19NHB}. Previous studies clearly show that, in Weibo, anger is more influential than joy~\cite{fan2014anger}, and in Twitter, negative content spreads faster than positive content~\cite{Ferra2015PJ}. However, several other studies have opposite conclusion, suggesting that positive emotions are more contagious online~\cite{coviello2014detecting,ferrara2015measuring,gruzd2011happiness}. Given these controversial conclusions, we propose that negative emotions may need a finer classification as emotions like anger, disgust and sadness are all negative but quite different in other dimensions such as arousal~\cite{wundt1907outlines}, which may also change the strength of contagion~\cite{jonah2011arousal,jonah2012what}. In this study, by finely splitting negative emotions into anger, disgust and sadness, we are the first to offer evidence that anger spreads faster than joy in social media such as Weibo. As written in an ancient Chinese book entitled ``Zhongyong (The Doctrine of the Mean)'' more than two thousand years ago, ``when the feelings (e.g. pleasure, anger, sadness, or joy) have been stirred, and they act in their due degree, there ensues what may be called the state of HARMONY.'' Our study implies that we should place a stronger emphasis on anger in both personal emotional management and in collective mood understanding to make social media reach this state.

In online social media platforms, it is much easier to create a connection than in offline scenarios, and the platform also encourage users to connect with more people. Therefore, online users tend to have larger networks and could be exposed to many weakly-connected friends~\cite{10.1145/1242572.1242685,GOLDENBERG2020316}. However, limited by Dunbar's number, users can only maintain a small amount of strong relationships~\cite{dunbarnumber,zhao-dunbar}. Previous study reveals that weak connections dominate Facebook~\cite{Meo2014}. In our follow graph, the percentage of reciprocity connections is only 8.3\%, indicating that in Weibo, weak ties also account for majority proportion. Meanwhile, our results demenstrate that the preference of weak ties speed up anger contagion, indicating that anger may spread much faster in online social media, which contain large amount of weak connections, than in real-world circumstances. In addition, weak ties are conventionally thought to be crucial in information spread~\cite{granovetter1973strength,UBHD2028615}, through which novel information like innovations can be spilled out and then widely disseminated. That's is to say, with the emergence and development of social media, the proliferation of weak ties could make both novel information and anger spread fast in modern societies, resulting in an unexpected dilemma of being well informed but easily frenzied. From this perspective, our results further suggest that the dominance of weak ties in social media could lead to a profound yet previously not realized trade-off between accelerating the flow of innovations and promoting the spread of anger. This trade-off, which is named as the \textit{Double-Edged Sword of Weak Ties} in this study, suggests that weak ties in social media should therefore be leveraged with the principle of a primed balance between novel ideas and irrational emotions. Specifically, for innovational information that will be widely passed around through weak ties, the emotions it carries needs a careful scrutiny to avoid the undesired contagion of anger.

The way anger is expressed and experienced on the Internet is gaining attention. Rant-site visitors have self-reported that posting angry messages produces immediate feelings of relaxation~\cite{martin2013anger}, which makes posting anger an effective method for self-regulating mood. However, considering the high contagion of anger, an increase in angry expressions on the Internet might arouse negative shifts in the mood of the crowds connected to the posters~\cite{martin2013anger,Park2012}. Moreover, anger's preference for weak ties make it more likely to spread to ``strangers'' on the Internet. Previous studies also show that anger improves the exchange of emotions~\cite{Crockett2017MoralOutrage,Zomeren12Protesters}, driving for many social movements such as Arab Spring~\cite{ghonim2012revolution}. Users who want to assuage their anger by posting on the Internet should be made to understand the possible impact that such posts can have on their online social networks. Even in offline scenarios such as the ``road rage''~\cite{roadrage} that occurs during rush hours in China, anger among strangers can spread quickly and cause aggressive driving or even accidents. We suggest that in personal anger management, the possibility of infecting strangers---however unexpected--should be seriously considered.

Online social media has become the most ubiquitous platform for collective intelligence, in which various signals generated by the massive numbers of connected individuals provide a foundation for understanding collective behavior. It is even possible that in the post-truth era, public opinion might be driven or shaped by emotional appeals rather than objective facts~\cite{posttruth}. However, how emotion contagion affects individual behavior in the aggregate---particularly individual intelligence---is rarely considered. In our findings, the spread of emotion, especially the high contagion and high velocity of anger, might have large implications concerning collective behavior in cyberspace, particularly with regard to crowdsourcing. It has even been stated that emotion such as anger can be a threat to reasoned argument~\cite{marsella2014computationally}. For example, outrage provoked in massive numbers of emotional individuals can profoundly bias public opinion, might originate with the fast and abroad spread of anger but not be due to the event itself. Meanwhile, the fast spread of anger also offers a new perspective for picturing the emotional behavior of crowds on the Internet. We suggest that in scenarios such as crowdsourcing and in understanding collective behavior, angry users should be treated carefully to negate or reduce their possible impact on the fair judgment of the observations in the big data era. In addition, reducing weak ties can function effectively in controlling the diffusion of Internet-fueled outrage and help make crowds more rational and smarter.

In contrast, happiness is believed to unify and cluster within communities~\cite{connected}. Our findings about joy's preferences for stronger ties supports this concept (see~Figs. \ref{fig:weak_tie} and \ref{fig:graph}), implying that the stronger ties inside communities are more suitable for disseminating joyful content in online social media. Meanwhile, self-reports from Facebook users also testify that communication with strong ties is associated with improvements in well-being~\cite{Burke2016}, which further indicates that our findings from the computational viewpoint are solid. 

\section*{Conclusion}
\label{sec:con}

Rather than using self-reports in controlled experiments, this study collected natural and emotional postings from the Weibo social network to investigate the mechanism of emotion contagion in detail from the new viewpoint of computational social science. For the first time, we offer solid evidence that anger spreads faster and wider than joy in social media because it disseminates preferentially through weak ties. Our findings shed light on both personal anger management and in understanding collective behavior.

This study inevitably has limitations. It is generally accepted that emotion expression is culture dependent and that demographics such as gender also matter~\cite{association_theory,hu2016ambivalence}, suggesting that exploring how anger spreads in Twitter and how male and female responses differ regarding emotion contagion may be of great significance in future work.



\begin{addendum}
 \item This work was supported by NSFC (Grant Nos. 71871006 and 61421003) and the fund of the State Key Lab of Software Development Environment (Grant No. SKLSDE-2019ZX-06).
 
 \item[Competing Interests] The authors declare that they have no
 competing financial interests.
\end{addendum}


\end{document}